\documentclass[aps,prb,superscriptaddress,showpacs,twocolumn]{revtex4}

\usepackage{psfig}

\input epsf

\def\prb{Phys. Rev. B }

\def\be{\begin{equation}}
\def\ee{\end{equation}}
\def\ba{\begin{eqnarray}}
\def\ea{\end{eqnarray}}

\def\124{YBa$_2$Cu$_4$O$_8$ }

\def\C60{A$_x$C$_{60}$ }

\begin{document}

\title
{Disorder Effects in Fluctuating One-Dimensional Interacting Systems}

\author{U. London}
\affiliation{Racah Institute of Physics, The Hebrew University,
Jerusalem 91904, Israel}

\author{T. Giamarchi}
\affiliation{University of Geneva, DPMC, 24 Quai Ernest Ansermet, CH-1211
Geneva 4, Switzerland}

\author{D. Orgad}
\affiliation{Racah Institute of Physics, The Hebrew University,
Jerusalem 91904, Israel}

\date{\today}

\begin{abstract}
The zero temperature localization of interacting electrons coupled
to a two-dimensional quenched random potential, and constrained to
move on a fluctuating one-dimensional string embedded in the
disordered plane, is studied using a perturbative renormalization
group approach. In the reference frame of the electrons the
impurities are dynamical and their localizing effect is expected to
decrease. We consider several models for the string dynamics and
find that while the extent of the delocalized regime indeed grows
with the degree of string fluctuations,
the critical interaction strength, which determines the
localization-delocalization transition for infinitesimal disorder,
does not change unless the fluctuations are softer than those of
a simple elastic string.
\end{abstract}


\maketitle

\section{Introduction}
\label{sec:intro}

Quenched disorder is a relevant perturbation in one- and
two-dimensional non-interacting fermionic systems\cite{LeeRam}. In
these low dimensions the resulting fully localized state can be
averted at zero temperature only under the influence of
interactions. Such interaction-induced delocalization has been
demonstrated in one-dimensional systems with strong
attraction\cite{Apel,gs}, and evidence for its potential existence
in two dimensions, even in the presence of repulsive forces, has
also been put forward\cite{Finkel,Castellani,Punn}. Most of our
knowledge of the interplay between disorder and interactions is for
static disorder, and much less is known about the interplay between
interactions and time-dependent disorder\cite{feldman,martin}.

Although for quantum systems, time-independent disorder is the 
standard case, one can encounter situations where the motion of the
electrons is constrained to fluctuating geometries which are embedded in a 
static disordered space. In such a case, to the electrons, the disorder 
appears to change in time with temporal correlations which are inherited 
from the dynamics of the fluctuating geometry. Relevant realizations 
of this scenario include excitations in the core of vortex states
\cite{caroli}, and the 
cuprate high-temperature superconductors, in which the importance of
self-organized, fluctuating quasi-one-dimensional electronic
structures within the (disordered) copper-oxide planes has been
pointed out\cite{stripereviews}. On the technical side, 
introducing the time dependence of the scattering potential in such
a way has the advantage of enabling us to treat the
problem using standard methods of equilibrium quantum statistical
mechanics. 

We therefore devote the present paper to the study of the consequences 
of coupling interacting electrons, constrained to move along a fluctuating 
string embedded in a plane, to static disorder within the plane.  
We consider several string Hamiltonians including that of a rigid string 
inside a harmonic well, a stretchable elastic string and a floppy string of
fixed length. These models form a hierarchy in terms of an
increasing degree of string fluctuations. Assuming weak disorder and
long wavelength fluctuations we investigate the relations between the
electron-electron interactions, disorder and string dynamics using a
renormalization group analysis. Our one-loop treatment of the
problem captures the mutual renormalization of these three elements.
It ignores subtle interference effects due to coherent scattering
from many impurities, but in one dimension (unlike in two or more
dimensions) where the localization length is known to be of the
order of the mean free path, such effects are not expected to play
an essential role. As anticipated, we find that the extent of the
delocalized phase increases when the string fluctuations become more
pronounced, as is demonstrated for example by Fig.~\ref{KDrigid}.
However, in the cases of the rigid and elastic strings, the critical
point that separates the localized and delocalized phases for
infinitesimal disorder remains at the same critical interaction
strength as for a static string. Only when the softer fluctuations
produced by the floppy string dynamics are considered, one finds
that the critical point shifts towards smaller values of
interactions.

The main body of the paper is composed of three sections, each
dealing with one of the realizations of the string dynamics as
indicated above. Every one of these sections introduces the model
for the coupled electronic and string degrees of freedom, in the
presence of disorder. The renormalization equations are then derived
and the resulting zero-temperature phase diagram discussed. The
effects of the forward-scattering part of the disorder on the
charge-density wave (CDW) and spin-density wave (SDW) correlations
are also given. Some of the details pertaining to the derivation of
the renormalization equations are relegated to the appendices.

\section{A Rigid String in a Parabolic Well}
\label{sec:Rigid}

\subsection{The Model}
\label{modelrigid}

We begin by considering interacting electrons which are constrained to move
on a straight rigid string of fixed length $L$, embedded in a two-dimensional
disordered plane. The string is assumed to oscillate inside a parabolic
potential and its state is characterized by its deviation $u(\tau)$ from
the classical equilibrium configuration at the bottom of the well (which we
define in the following as the $x$-axis). The
string Lagrangian in imaginary time is therefore
\be
\label{stringLrigid}
L_s=\frac{M}{2}(\partial_\tau u)^2+\frac{M\omega_0^2}{2}u^2 \; ,
\ee
where $M$ is the mass of the string and $\omega_0$ its oscillation
frequency.

The low-energy physics of the one-dimensional electron gas (1DEG) is
captured by the Tomonaga-Luttinger model\cite{1dreview} in which the
spectrum is linearized around the two Fermi points at $\pm k_F$, and
the electronic operator
$\psi_{\sigma=\{\uparrow,\downarrow\}}$ is decomposed in terms of slowly
varying left ($\nu=-$) and right ($\nu=+$) moving components:
$\psi_\sigma(x)=e^{-ik_F x}\psi_{-,\sigma}(x)+e^{ik_F x}\psi_{+,\sigma}(x)$.
We follow the standard notation and parameterize the interactions between the
electrons according to
$\frac{1}{2} \sum_{\nu,\sigma,\sigma'}
[g_4\rho_{\nu,\sigma}\rho_{\nu,\sigma'}
+g_2\rho_{\nu,\sigma}\rho_{-\nu,\sigma'}
-g_{1\parallel}\delta_{\sigma,\sigma'}\rho_{\nu,\sigma}\rho_{-\nu,\sigma'}
+g_{1\perp}\delta_{\sigma,-\sigma'}\psi^{\dagger}_{\nu,\sigma}
\psi^{\dagger}_{-\nu,\sigma'}\psi_{\nu,\sigma'}\psi_{-\nu,\sigma}]$,
where $\rho_{\nu,\sigma}=\psi^{\dagger}_{\nu,\sigma}\psi_{\nu,\sigma}$.
The model can also be expressed in terms of charge and spin bosonic
fields $\phi_{\alpha=\{c,s\}}$ and their conjugated momenta
$\partial_x\theta_\alpha$, in terms of which the electronic Lagrangian
is given by (here and throughout $\hbar=1$)
\ba
\label{electLrigid}
\nonumber
L_e=&&\int_0^{L} dx \sum_{\alpha=c,s}
\bigg[-i\partial_\tau\phi_\alpha\partial_x\theta_\alpha
+\frac{{v}_\alpha {K}_\alpha}{2}(\partial_x\theta_\alpha)^2 \\
&&+\frac{{v}_\alpha}{2{K}_\alpha}(\partial_x\phi_\alpha)^2\bigg]
+\frac{g_{1\perp}}{2(\pi a)^2}\cos(\sqrt{8\pi}\phi_s) \; ,
\ea
where $a$ is a short-distance cutoff of the order of the lattice constant,
and the velocities and Luttinger parameters are related to the Fermi
velocity $v_F$ and the interaction couplings according to
\ba
\label{vK}
\nonumber
v_c&=&\frac{1}{2\pi}\sqrt{(2\pi v_F+2g_4)^2-(2g_2-g_{1\parallel})^2} \; , \\
\nonumber
v_s&=&\frac{1}{2\pi}\sqrt{(2\pi v_F)^2-g_{1\parallel}^2} \; , \\
K_c&=&\sqrt{\frac{2\pi v_F +2g_4-2g_2+g_{1\parallel}}
{2\pi v_F +2g_4+2g_2-g_{1\parallel}}} \; , \\
\nonumber
K_s&=&\sqrt{\frac{2\pi v_F +g_{1\parallel}}
{2\pi v_F -g_{1\parallel}}} \; .
\ea

As can be seen from our analysis of the more general problem of a
fluctuating string, as presented in Sec.~\ref{sec:Gaussian}, in the
present model (which is equivalent to a 1DEG coupled to an optical
phonon in the transverse direction) the only coupling between the
electrons and the string (in the absence of disorder) is via a
renormalization of the string mass. We assume that this
renormalization  $M\rightarrow M+mN_e$, where $m$ is the electronic
mass and $N_e$ is the number of electrons on the string, has been
incorporated into the definition of the parameters which appear in
the string Lagrangian, Eq.~(\ref{stringLrigid}).

The plane containing the string and the electrons is taken to include
a weak random potential which couples to the electrons, but not directly
to the string. This choice is motivated primarily by the situation in
the cuprates where the ``strings'' are actually defined as the loci of
points occupied by holes in the plane, and as such couple to the
disorder only indirectly via the electronic degrees of freedom.
The forward scattering $k\sim 0$ and the backward
scattering $k\sim 2k_F$ components of the impurities potential are assumed
to be uncorrelated Gaussian random fields $\eta$ and $\xi$ with
$\overline{\eta(x,y)\eta(x',y')}=D_f\delta(x-x')\delta(y-y')$, and
$\overline{\xi^*(x,y)\xi(x',y')}=D_b\delta(x-x')\delta(y-y')$,
respectively\cite{gs}. The coupling of the electrons to the impurities
is given by
\ba
\label{forwardLrigid}
\nonumber
L_{dis}^f&=&-\sum_{\nu,\sigma}\int_0^L dx\, \eta[x,u(\tau)]
\rho_{\nu,\sigma}(x,\tau) \\
&=&-\int_0^L dx \sqrt{\frac{2}{\pi}}\eta[x,u(\tau)]\partial_x\phi_c(x,\tau)
\; ,
\ea
and
\ba
\label{backwardLrigid}
\nonumber
L_{dis}^b&=&-\sum_{\sigma}\int_0^L dx \,  \xi[x,u(\tau)]
\psi_{-,\sigma}^\dagger(x,\tau)\psi_{+,\sigma}(x,\tau) \\
\nonumber
&&+{\rm H.c.} \\
\nonumber
&=&-\int_0^L \frac{dx}{\pi a} \xi[x,u(\tau)]e^{i\sqrt{2\pi}\phi_c(x,\tau)}
\cos[\sqrt{2\pi}\phi_s(x,\tau)] \\
&&+{\rm H.c.} \ea 
Here the constraint which confines the electrons
to the string manifests itself in the position at which the disorder
potential is evaluated. This, in turn, induces a coupling between
the electronic and string degrees of freedom, and leads, in the
reference frame of the electrons, to an effective time-dependent
disorder. This time dependence precludes the possibility of gauging
out the forward scattering disorder component from the action by an
appropriate shift of $\phi_c$, as was done in Ref.~\onlinecite{gs}.
Instead, both the forward scattering and the backward scattering
components will be treated on equal footing within one-loop
renormalization group analysis.

To this end, we use the replica trick and average the replicated
action over the random fields $\eta$ and $\xi$. We will obtain the
renormalization equations to first order in $D_b$ and $D_f$, and to
second order in $g_{1\perp}$, which is also assumed small. To this
order, the replica indices play no role and consequently they will
be omitted in the following. The averaged action is therefore
\begin{widetext}
\ba
\label{averagedS}
\nonumber
S&=&\int d\tau \left[L_s+L_e\right] - \frac{D_f}{2}\int d\tau d\tau' dx \,
\delta[u(\tau)-u(\tau')]\sum_{\nu,\nu',\sigma,\sigma'}
\rho_{\nu,\sigma}(x,\tau)\rho_{\nu',\sigma'}(x,\tau') \\
&-&\frac{D_b}{2}\int d\tau d\tau' dx \, \delta[u(\tau)-u(\tau')]
\sum_{\sigma,\sigma'}\left[\psi_{+,\sigma}^\dagger(x,\tau)\psi_{-,\sigma}
(x,\tau)\psi_{-,\sigma'}^\dagger(x,\tau')\psi_{+,\sigma'}(x,\tau')
+{\rm H.c.}\right] \; . 
\ea
\end{widetext}

\subsection{Renormalization and Phase Diagram}
\label{renormrigid}

We derive the renormalization flow equations for the model by
requiring invariance of the long-wavelength low-frequency behavior
of the correlation functions under a change of the
cutoff\cite{gs,jose}. We begin with the electronic correlations and
integrate out the string degrees of freedom in order to obtain an
effective electronic action. The result is an action which is
derived from Eq.~(\ref{averagedS}) by replacing
$\delta[u(\tau)-u(\tau')]$ by its average over the string dynamics
\ba \label{F}
\nonumber
F(\tau-\tau')&\equiv&\langle \delta[u(\tau)-u(\tau')] \rangle \\
\nonumber
&=& \frac{1}{2\pi}\int_{-\infty}^\infty d\lambda \langle
e^{i\lambda[u(\tau)-u(\tau')]}\rangle \\
&=&\frac{1}{\sqrt{2\pi}}
\langle\left[u(\tau)-u(\tau')\right]^2\rangle^{-\frac{1}{2}} \; .
\ea
As we demonstrate below, it is this function which determines the
modification in the renormalization flow of the model as a result of the
string fluctuations. In the present model one finds
\be
\label{Frigid}
F(\tau)=\frac{1}{2\sqrt{\pi}\lambda}\frac{1}{\sqrt{1-e^{-\omega_0|\tau|}}}
\; ,
\ee
where the length
\be
\label{lambdarigid}
\lambda=\frac{1}{\sqrt{2M\omega_0}} \; ,
\ee
is a measure of the amplitude of the string fluctuations inside the
harmonic well.

Our renormalization procedure is akin to real-space renormalization
in the sense that electronic degrees of freedom at points $(x,\tau)$
and $(x,\tau')$ are identified inside the interval $|\tau-\tau'|<a/v_s$.
Upon such identification, the $|\tau-\tau'|<a/v_s$ part of the
forward scattering term in Eq. (\ref{averagedS}) is equivalent to $g_2$
and $g_4$ interaction processes, while the corresponding part of the backward
scattering term is equivalent to $g_{1\parallel}$ and $g_{1\perp}$
processes. As a result one can absorb the contributions of these parts
via a redefinition of the interaction parameters according to
\ba
\label{barg}
\nonumber
\overline{g}_{2,4} &=& g_{2,4} -2D_f \int_0^{a/v_s} d\tau F(\tau) \; , \\
\overline{g}_{1\parallel,1\perp} &=& g_{1\parallel,1\perp}
-2D_b \int_0^{a/v_s} d\tau F(\tau) \; .
\ea
The effective electronic action then reads

\begin{widetext}

\ba
\label{effectiveS}
\overline{S}_e&=&\int d\tau \overline{L}_e - \frac{D_f}{\pi}
\int_{|\tau-\tau'|>a/v_s} d\tau d\tau' dx \,
F(\tau-\tau')\partial_x\phi_c(x,\tau)\partial_{x'}\phi_c(x,\tau')  \\
\nonumber
&-&\frac{D_b}{(\pi a)^2}\int_{|\tau-\tau'|>a/v_s} d\tau d\tau' dx
\,F(\tau-\tau')
\cos[\sqrt{2\pi}\phi_s(x,\tau)]\cos[\sqrt{2\pi}\phi_s(x,\tau')]
\cos[\sqrt{2\pi}\phi_c(x,\tau)-\sqrt{2\pi}\phi_c(x,\tau')] \; , 
\ea
\end{widetext}
where $\overline{L}_e$ has the same form as Eq. (\ref{electLrigid}),
but with modified velocities and Luttinger parameters, which to
first order in the disorder strength are given by \ba \label{barvK}
\nonumber \overline{K}_c\!\!&=&\!\! K_c+\left[\frac{2K_c^2}{\pi v_c}D_f
-\frac{1}{2\pi v_c}(1+K_c^2)D_b\right]\!
\int_0^{a/v_s} \!\! d\tau F(\tau)  , \\
\nonumber
\overline{K}_s\!\!&=&\!\! K_s-\frac{1}{2\pi v_s}(1+K_s^2)D_b
\int_0^{a/v_s} \!\! d\tau F(\tau)  , \\
\nonumber
\overline{v}_c\!\!&=&\!\!v_c-\left[\frac{2K_c}{\pi}D_f
+\frac{1}{2\pi}\!\left(\frac{1}{K_c}-K_c\!\right)\!D_b\right]\!
\int_0^{a/v_s} \!\! d\tau F(\tau)  , \\
\nonumber
\overline{v}_s\!\!&=&\!\!v_s-\frac{1}{2\pi}\!\left(\frac{1}{K_s}-K_s\!\right)
\!D_b\int_0^{a/v_s} \!\! d\tau F(\tau)  , \\
\nonumber
\overline{y}\!&=&\!y+\left[\frac{y}{2\pi v_s}\left(\frac{1}{K_s}-K_s\right)
-\frac{2}{\pi v_s}\right]\!D_b
\int_0^{a/v_s} \!\! d\tau F(\tau)  , \\
\ea
as a consequence of Eqs. (\ref{vK},\ref{barg}). Here and in the
following we have defined the dimensionless spin-flipping backward
scattering coupling
\be
y=\frac{g_{1\perp}}{\pi v_s} \; .
\ee

The derivation of the renormalization equations for the parameters
which appear in the effective electronic action, Eq.
(\ref{effectiveS}), follows closely Ref. \onlinecite{gs}. We give
some details concerning the contribution of the forward scattering
disorder to these equations in Appendix \ref{forwardRG}, where the
latter are also listed. We then use them, together with Eqs.
(\ref{Frigid},\ref{barvK}), to obtain the renormalization equations
of the electronic and disorder parameters in the original disordered
averaged action, Eq. (\ref{averagedS}). The result is
\begin{widetext}
\ba
\label{RGrigid}
\nonumber
\frac{dK_c}{d\ell}&=&-\frac{1}{2}\frac{v_c}{v_s}\frac{{\cal D}_b}
{\sqrt{1-e^{-\varpi}}}\left[K_c^2-\frac{1}{2}
\left(\frac{v_c}{v_s}\right)^{K_c-2}(1+K_c^2)\right]\\
\nonumber
&+&\frac{{\cal D}_b}{4}
\left(\frac{v_c}{v_s}\right)^{K_c-1}(2-K_c-K_s-y)(1+K_c^2)
\left[1+\frac{2}{\varpi}\ln\left(1+\sqrt{1-e^{-\varpi}}\right)\right] \; ,\\
\nonumber
\frac{dv_c}{d\ell}&=&-\frac{1}{2}\frac{v_c^2}{v_s}\frac{{\cal D}_b}
{\sqrt{1-e^{-\varpi}}}\left[K_c-\frac{1}{2}
\left(\frac{v_c}{v_s}\right)^{K_c-2}\left(\frac{1}{K_c}-K_c\right)\right]\\
\nonumber
&+&\frac{{\cal D}_b}{4}
\left(\frac{v_c}{v_s}\right)^{K_c-1}(2-K_c-K_s-y)\left(\frac{1}{K_c}-K_c\right)
\left[1+\frac{2}{\varpi}\ln\left(1+\sqrt{1-e^{-\varpi}}\right)\right] \; ,\\
\nonumber
\frac{dK_s}{d\ell}&=&-\frac{1}{2}y^2 K_s^2-\frac{1}{2}\frac{{\cal D}_b}
{\sqrt{1-e^{-\varpi}}}\left[K_s^2-\frac{1}{2}
\left(\frac{v_c}{v_s}\right)^{K_c}(1+K_s^2)\right]\\
\nonumber
&+&\frac{{\cal D}_b}{4}
\left(\frac{v_c}{v_s}\right)^{K_c}\left[(2-K_c-K_s-y)(1+K_s^2)+4K_s^2y\right]
\left[1+\frac{2}{\varpi}\ln\left(1+\sqrt{1-e^{-\varpi}}\right)\right] \; ,\\
\frac{dv_s}{d\ell}&=&-\frac{1}{2}\frac{{\cal D}_b}
{\sqrt{1-e^{-\varpi}}}\left[K_s-\frac{1}{2}
\left(\frac{v_c}{v_s}\right)^{K_c}\left(\frac{1}{K_s}-K_s\right)\right]\\
\nonumber
&+&\frac{{\cal D}_b}{4}
\left(\frac{v_c}{v_s}\right)^{K_c}(2-K_c-K_s-y)\left(\frac{1}{K_s}-K_s\right)
\left[1+\frac{2}{\varpi}\ln\left(1+\sqrt{1-e^{-\varpi}}\right)\right] \; ,\\
\nonumber
\frac{dy}{d\ell}&=&(2-2K_s)y-\frac{{\cal D}_b}
{\sqrt{1-e^{-\varpi}}}\left\{1-\left(\frac{v_c}{v_s}\right)^{K_c}\left[
1-\frac{y}{4}\left(\frac{1}{K_s}-K_s\right)\right]\right\} \\
\nonumber
&+&\frac{{\cal D}_b}{4}
\left(\frac{v_c}{v_s}\right)^{K_c}
\left\{4(K_s-K_c)-y\left[1-K_s^2+(2-K_c)\left(\frac{1}{K_s}-K_s\right)
\right]\right\}
\left[1+\frac{2}{\varpi}\ln\left(1+\sqrt{1-e^{-\varpi}}\right)\right] \; ,\\
\nonumber
\frac{d{\cal D}_f}{d\ell}&=&{\cal D}_f \; , \\
\nonumber
\frac{d{\cal D}_b}{d\ell}&=&(3-K_c-K_s-y){\cal D}_b \; ,
\ea
\end{widetext}
where the running cutoff is given by $a=a_0e^\ell$ and where we have
defined the dimensionless quantities \ba \label{D} \nonumber {\cal
D}_b&=&\frac{D_b a}{\pi^{3/2} v_s^2 \lambda}
\left(\frac{v_s}{v_c}\right)^{K_c} \; , \\
{\cal D}_f&=&\frac{D_f a}{\pi^{3/2} v_c^2 \lambda} \; , \\
\nonumber
\varpi&=&\frac{a\omega_0}{v_s} \; .
\ea

Expressing the flow equations in terms of the original velocities and
Luttinger parameters has the advantage that these, unlike the barred
quantities of Eq.~(\ref{barvK}), are related to the electronic interaction
couplings in a familiar manner. As a result, it is straightforward to
check that if the system is initially at the non-interacting point
$K_c=K_s=1$, $v_c=v_s=v_F$ and $y=0$, then it stays there in the course
of the renormalization, i.e. $dK_c/d\ell=dK_s/d\ell=dy/d\ell=0$. In
other words, a system of independent electrons remains so even in the
presence of (time-dependent) disorder. Secondly, the equations preserve
spin-rotation symmetry. If one starts from a spin-rotation invariant
Hamiltonian $(g_{1\parallel}=g_{1\perp})$ it continues to respect this
symmetry during the renormalization process. This fact can be easily checked
for small $g_1$, in which case $K_s\approx 1+y/2$. The flow maintains this
relation\cite{invcomm} since it satisfies $dK_s/d\ell=1/2 (dy/d\ell)$.

Note that the renormalization equations for the electronic parameters do
not include the impurity forward scattering amplitude ${\cal D}_f$. It
does appear in the flow equations, Eq.~(\ref{RGbar}), for the parameters
in the effective action $\overline{S}_e$, but cancels out for the original
parameters as a consequence of the relations given in Eq.~(\ref{barvK}).
This fact has been demonstrated in the case of a static string\cite{gs},
where it is possible to completely absorb the forward scattering due to
impurities by a redefinition of the field $\phi_c$. Here we show how it
also arises when one treats the scattering perturbatively and find that it
extends also to cases where the string is dynamical.

\begin{figure}[h!!!]
\setlength{\unitlength}{1in}
\begin{picture}(3.2,3.3)(0,-2.1)
\put(-0.08,-1.4){\psfig{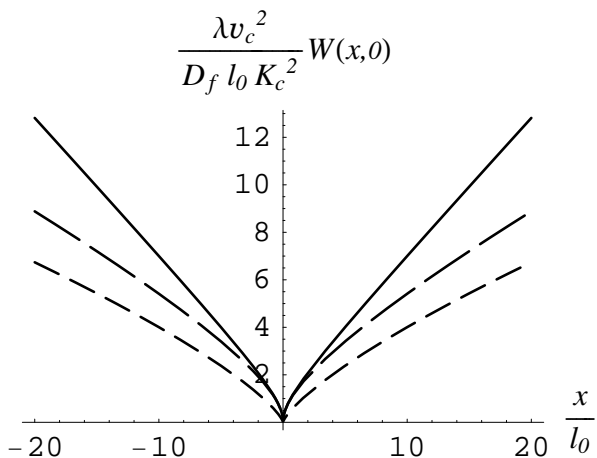}}
\end{picture}
\vspace{-1.8cm}
\caption{The function $W(x,0)$ for the different models: a rigid string in a
parabolic well (solid line), an elastic string (long-dashed line), and a
floppy string (dashed line). Both axes are scaled by $l_0=v_c/\omega_0$,
$(v_c/v_u)a$, and $4\pi (v_c/\gamma)\lambda$, in the three cases respectively.}
\label{Wfig}
\end{figure}

The forward scattering does influence, however, various correlation
functions, most notably those of the charge density wave $R_{\rm CDW}$
and spin density wave $R_{\rm SDW}$ order parameters, which are both
proportional to the function $R_c$ defined in Appendix~\ref{forwardRG}.
Consequently one finds $R_{\rm CDW,SDW}(x,\tau)\propto e^{-W(x,\tau)}$,
with the exact result\cite{gs} for the static string
$W(x,\tau)=2D^{\rm 1d}_f(K_c/v_c)^2|x|$ (here $ D^{\rm 1d}_f$
measures the disorder correlations along one dimension and as such
differs by factor $(2\sqrt{\pi}\lambda)^{-1}$ from $D_f$ which appears in
our analysis.) When fluctuations are included we are able to calculate,
using Eq. (\ref{fscontrib2}), the asymptotic
behavior of $W$ to first order in the disorder and obtain
\be
\label{Wxrigid}
W(x,0)=\left\{
\begin{array}{l@{\quad : \quad}l}
\frac{3\pi}{\sqrt{2}}\frac{{v_c \cal D}_f}{v_s \varpi}K_c^2
\sqrt{\frac{\omega_0 x}{v_c}}+{\cal O}(x^2) & \!\omega_0|x|\ll v_c \\
\pi{\cal D}_f K_c^2\frac{|x|}{a}+{\cal O}(\ln|x|) & \!\omega_0|x|\gg v_c
\end{array}
\right. ,
\ee
and
\be
\label{Wtrigid}
W(0,\tau)=\left\{
\begin{array}{l@{\quad : \quad}l}
\pi\frac{{v_c \cal D}_f}{v_s \varpi}K_c^2\sqrt{\omega_0|\tau|}+
{\cal O}(\tau^2)\!\! & \!\!\!\omega_0|\tau|\ll 1 \\
\ln 4\frac{{v_c \cal D}_f}{v_s \varpi}K_c^2\ln(\omega_0|\tau|)
+{\cal O}(1)\!\! & \!\!\!\omega_0|\tau|\gg 1
\end{array}
\right. .
\ee

The full behavior of $W(x,0)$ is presented in Fig. \ref{Wfig}. The
exponential suppression, at large distances, of the charge and spin density
wave correlations due to forward scattering is similar to that in the
case of a static string. Note that the static limit result is obtained
by taking $\omega_0\rightarrow\infty$ in order to keep the string in its
ground state, at least as far as the low-energy electronic physics is
concerned. Since ${\cal D}_f\propto \lambda^{-1}$ the latter
limit implies the need to scale $D_f$ with $\lambda$ in order to remain in
the perturbative regime, where our results are valid.
In contrast to the case of the static string the fluctuations also increase 
the algebraic decay of $R_{\rm CDW}$ and $R_{\rm SDW}$ in the time domain 
and lead to a decrease in the conductivity and pairing correlations. 
We will present results pertaining to this latter issue elsewhere.

In order to derive the flow equations of the string parameters we first
average the action, Eq. (\ref{averagedS}), over the electronic degrees of
freedom. In accordance with the procedure we have utilized previously,
here too, we distinguish between electronic degrees of freedom only as
long as their time separation is larger than $a/v_s$. Consequently, for
shorter separations the electronic operators which appear in the forward
and backward scattering parts of the action, Eq. (\ref{averagedS}), become
$(\partial_x\phi_c)^2$ and $[(\partial_x\theta_c)^2-(\partial_x\phi_c)^2
+(\partial_x\theta_s)^2-(\partial_x\phi_s)^2]/4\pi+1/(2\pi^2 a^2)
\cos(\sqrt{8\pi}\phi_s)$, respectively. The averages of these operators
diverge quadratically and need to be regularized using an appropriate
cutoff (except for the cosine operator which vanishes whenever the system
is in a spin-gapless phase). To do so, we note that in the course of the
renormalization procedure the short time piece of the integral is being
constructed by adding infinitesimal slices from the initial cutoff $a_0$ to
the its running value $a$. We therefore evaluate the electronic averages
within each slice using the value of the cutoff at the time the slice was
added along the renormalization process. The outcome is the following
effective action for the string degrees of freedom
\begin{widetext}

\ba
\label{effSs}
\nonumber
\overline{S}_s&=&\int d\tau S_s
-L\left\{\frac{D_f K_c}{2(\pi v_c)^2}+\frac{D_b}{8\pi^2}\left[\frac{1}{v_c^2}
\left(\frac{1}{K_c}-K_c\right)+\frac{1}{v_s^2}\left(\frac{1}{K_s}-K_s\right)
\right]\right\}\int_{a/v_s>|\tau-\tau'|>a_0/v_s}d\tau d\tau' \,
\frac{\delta[u(\tau)-u(\tau')]}{(\tau-\tau')^2} \\
&-&L\int_{|\tau-\tau'|>a/v_s}d\tau d\tau' \, \delta[u(\tau)-u(\tau')]
\left[\frac{D_f}{2\pi^2}\frac{K_c}{v_c^2(\tau-\tau')^2}
+\frac{D_b}{2(\pi a)^2}\left(\frac{a}{v_c|\tau-\tau'|}\right)^{K_c}
\left(\frac{a}{v_s|\tau-\tau'|}\right)^{K_s}\right] \; .
\ea

\end{widetext}
In Appendix \ref{stringRG} we use this action to derive the flow
equations for $\varpi$ and $\lambda$. Using the dimensionless string
length \be \label{Ld} {\cal L}=\frac{L}{2\pi a} \; , \ee the result
is
\begin{widetext}

\ba
\label{RGstrigid}
\nonumber
\frac{d\varpi}{d\ell}&=&\varpi+{\cal D}_b{\cal L}\varpi\left[[1-f(v,K)]
\frac{1-\cosh\varpi}{\varpi(1-e^{-\varpi})^{3/2}}
-f(v,K)(2-K_c-K_s-y)\int_{e^{-\ell}\varpi}^{\varpi}
d\eta \frac{1-\cosh\eta}{\eta^2(1-e^{-\eta})^{3/2}}\right] \; , \\
\nonumber
\frac{d\lambda}{d\ell}&=&-\frac{{\cal D}_b{\cal L}\lambda}{2}
\left[[1-f(v,K)]\frac{1-\cosh\varpi+\varpi\sinh\varpi}
{\varpi(1-e^{-\varpi})^{3/2}}
-f(v,K)(2-K_c-K_s-y)\int_{e^{-\ell}\varpi}^{\varpi} d\eta
\frac{1-\cosh\eta+\eta\sinh\eta}{\eta^2(1-e^{-\eta})^{3/2}}\right]\; , \\
\frac{d{\cal L}}{d\ell}&=&-{\cal L} \; ,
\ea

\end{widetext}
where we have introduced the compact notation \be \label{fvK}
f(v,K)\equiv\frac{1}{4}\left(\frac{v_c}{v_s}\right)^{K_c}\!\left[
\left(\frac{v_s}{v_c}\right)^2\!\left(\frac{1}{K_c}-K_c\right)+\frac{1}{K_s}
-K_s\right] . \ee

\begin{figure}[h!!!]
\setlength{\unitlength}{1in}
\begin{picture}(3.2,3.3)(0,-2.46)
\put(-0.08,-1.4){\psfig{figure=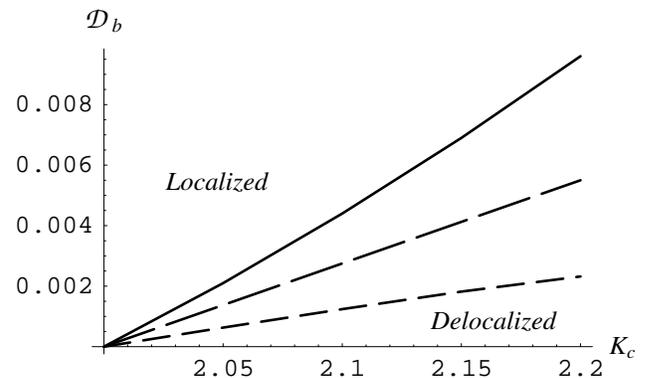,width=3.3in}}
\end{picture}
\vspace{-2.6cm}
\caption{The separatrix between the localized and delocalized phases in
the $K_c-{\cal D}_b$ plane for different values of the string fluctuation
frequency: $\varpi=1$ (solid line),  $\varpi=0.1$ (long-dashed line), and
$\varpi=0.01$ (dashed line). Here $v_c=v_F=1$, $y=0.1$, 
$K_s=\sqrt{(1+y/2)/(1-y/2)}$, $v_s=\sqrt{1-(y/2)^2}$, ${\cal L}=10$. 
For infinitesimal disorder the critical point separating the localized and 
delocalized regimes is identical to the static one. However, for finite 
disorder the size of the delocalized region is increased by the string
fluctuations.}
\label{KDrigid}
\end{figure}

Eqs. (\ref{RGrigid}) and (\ref{RGstrigid}) describe the interplay between
the electronic interactions, disorder and string dynamics. In the present
model the critical point which separates the regimes of relevant and
irrelevant disorder, in the limit of extremely weak bare impurity potential
${\cal D}_b\rightarrow 0$, is insensitive to the string fluctuations and is
given by the condition
\be
\label{criticalcond}
3-K_c-K_s^*=0 \; ,
\ee
where $K_s^*$ is the renormalized value of $K_s$ due to the coupled flow with
$y$. However, for finite disorder, fluctuation induced delocalizing
effects are present and are manifested in two ways. First, since
${\cal D}_b\propto \lambda^{-1}$, see Eq. (\ref{D}), large amplitude
oscillations tend to decrease the initial value of the effective disorder
strength. Secondly, the oscillations frequency defines a crossover scale
$\varpi\sim 1$, below which the disorder is more effective in renormalizing
the interactions towards more repulsive values, as can be seen from the
flow equation for $K_c$. Since strong attractive interactions are
necessary to drive the system into a delocalized phase, decreasing the
oscillation frequency makes this eventuality less likely to occur.
The combined outcome of these two mechanisms is an increase in the extent
of the delocalized region as one increases the oscillation amplitude and
frequency. The effect of varying the latter is presented in Fig. \ref{KDrigid}
where we plot the separatrix between the regimes of relevant and irrelevant
disorder for several oscillation frequencies while keeping $\lambda$ fixed.

The analysis of the various flow regimes, as determined by the
renormalization equations, is similar to the one presented in 
Ref.~\onlinecite{gs},
and we comment on it briefly. When both ${\cal
D}_b$ and $y$ flow to zero the system approaches a fixed line
parametrized by $K_c^*\geq 2$ (in the spin-rotationally invariant
case). This line corresponds to a delocalized phase with dominant
triplet pairing correlations. When $y$ is initially small or
negative it flows to large negative values and pins the spin field
$\phi_s$. At the same time ${\cal D}_b$ may still be irrelevant
provided its bare value is small and $K_c$ is large. For
infinitesimal disorder the latter requirement becomes $K_c>3$,
implying that the resulting delocalized phase is dominated by
singlet pairing fluctuations. In both cases the string parameters
$\omega_0$ and $\lambda$ decrease but attain finite fixed values
corresponding to a still fluctuating string.

When ${\cal D}_b$ scales to infinity the perturbative renormalization
equations can no longer be trusted. This strong coupling regime contains
the non-interacting system, which in the static case is known to be
localized, based on exact calculations. As long as one assumes that
there is no additional fixed point at intermediate coupling this fact
implies that the entire strong coupling phase is localized. For the
fluctuating string no such exact solutions are available. However,
the non-interacting problem is equivalent to that of a particle (with
anisotropic mass) moving inside a two-dimensional disordered potential
and a harmonic well in the $y$ direction. It's wavefunction, $\Psi(x,u)$,
is believed to be localized for any strength of the disorder, implying
localization  of the electrons and pinning of the string. The latter effect
is also reflected in the renormalization equations for the string parameters,
which in the case of relevant backward electronic scattering, flow towards
$\omega_0\rightarrow 0$ and $\lambda\rightarrow 0$.
The nature of the localized state depends on the sign of the renormalized
coupling $g_1^*$. The system is expected\cite{gs} to consist of localized
pairs of spins for $g_1^*<0$ and of isolated spins interacting via
antiferromagnetic superexchange when $g_1^*>0$.

\section{Elastic String}
\label{sec:Gaussian}

\subsection{The Model}
\label{modelGaussian}

Here we consider an elastic string whose projection on the $x$ axis is
of fixed length $L_x$, and which obeys periodic boundary conditions in
this direction. We consider the limit of a stiff and massive string
such that we can ignore overhangs and describe its state by
the deviation $u(x,\tau)$ from the classical equilibrium configuration.
The string dynamics is governed by the Lagrangian
\be
\label{stringLgaussian}
L_s=\int_0^{L_x} dx \left[\frac{\rho}{2}(\partial_\tau u)^2+
\frac{\sigma}{2}(\partial_x u)^2\right] \; ,
\ee
in which it is characterized by its linear mass density $\rho$
and tension $\sigma$. Alternatively, the string can be described
by the velocity of the elastic waves which it carries
\be
\label{vu}
v_u=\sqrt{\frac{\sigma}{\rho}} \;,
\ee
and by the length
\be
\label{lambdaG}
\lambda=\sqrt{\frac{v_u}{2\pi\sigma}} \; .
\ee
The latter is related to the average slope of the string, relative to its
equilibrium configuration, according to
\be
\label{lambdabar}
\sqrt{\langle(\partial_x u)^2\rangle}=\lambda/a \; .
\ee
The condition of a stiff and massive string implies $\lambda\ll a$.

In a previous publication, Ref. \onlinecite{string}, one of us has
demonstrated that the constraint which forces the particles to
remain on the string induces a non-trivial metric for the electrons,
and couples them to effective gauge potentials, which are functions
of the string degrees of freedom. In the absence of external
electromagnetic fields the metric and gauge potentials are given by
\ba \label{gaugepot} \nonumber
g(x,\tau)&=& 1+(\partial_x u)^2 \; , \\
{\cal A}_0(x,\tau)&=&-\frac{m}{2e}(\partial_\tau u)^2 \; , \\
\nonumber
{\cal A}_1(x,\tau)&=&-\frac{imc}{e}\partial_x u\partial_\tau u \; .
\ea
Assuming a fixed number of electrons, $N_e$, on the string we require that
their projected density $n_x=N_e/L_x$, or equivalently, their projected
Fermi wave-vector $k_{Fx}=\pi n_x/2$ obey $k_{Fx} a\gtrsim 1$,
such that the $2k_{Fx}$ components of the gauge potentials and the metric
may be neglected. We also assume that the energy associated with the short
wavelength string waves obeys $v_u/a\ll E_{Fx}=k_{Fx}^2/2m=
v_{Fx}k_{Fx}/2$.
Consequently, we disregard backward scattering due to string fluctuations.
Under these conditions the electronic Lagrangian is
\begin{widetext}

\ba
\label{Lefluct}
\nonumber
L_e=\int_0^{L_x} dx \Bigg\{&&
\sum_{\alpha=c,s}\left[-i\partial_\tau\phi_\alpha\partial_x\theta_\alpha
+\frac{\widetilde{v}_\alpha \widetilde{K}_\alpha}{2}
(\partial_x\theta_\alpha)^2+\frac{\widetilde{v}_\alpha}
{2\widetilde{K}_\alpha}(\partial_x\phi_\alpha)^2\right]
+g^{-\frac{1}{2}}\frac{g_{1\perp}}{2(\pi a)^2}\cos(\sqrt{8\pi}\phi_s) \\
-&&\left[e{\cal A}_0-g^{-1}E_{Fx}-\frac{1}{32m}g^{-3}(\partial_x g)^2\right]
\left( n_x + \sqrt{\frac{2}{\pi}} \partial_x\phi_c \right)
-\widetilde{v}_c\widetilde{K}_c\sqrt{\frac{2}{\pi}}\frac{e}{c}{\cal A}_1
\partial_x\theta_c +\widetilde{v}_c\widetilde{K}_c\frac{e^2}
{\pi c^2}{\cal A}_1^2\Bigg\} \, .
\ea
\end{widetext}
Here $\widetilde{v}_\alpha$ and $\widetilde{K}_\alpha$ are
derived from the previously defined quantities in Eq. (\ref{vK}) via
the replacements $v_F\rightarrow g^{-1}v_{Fx}$ and
$g_{1\parallel,2,4}\rightarrow g^{-1/2}g_{1\parallel,2,4}$ [this is
a result of assuming short range interactions of the form
$V(s-s')\propto\delta(s-s')$, where $s$ is the arc-length parameter
along the string].

The terms involving $n_x$ in Eq. (\ref{Lefluct}) induce a renormalization
of the bare string parameters at tree level. The first of them is identically
$mn_x\int_0^{L_x} dx (\partial_\tau u)^2/2$ and expresses the fact the
electrons are dragged with the string as it moves, thereby increasing its
effective mass density
\be
\label{rhorenorm}
\rho\rightarrow \rho+mn_x \; .
\ee
When expanded to second order in derivatives of $u$, the second term reads
$E_{Fx} n_x-2E_{Fx}n_x\int_0^{L_x} dx (\partial_x u)^2/2$.
It encodes the fact that when the number of electrons is fixed, their
total kinetic energy is lowered as the string fluctuates and becomes
longer on the average. This gain in kinetic energy favors a more flexible
string through the renormalization of its tension according to
\be
\label{sigmarenorm}
\sigma\rightarrow\sigma-2E_{Fx}n_x \; .
\ee
In the following we assume that these effects are small and have already
been incorporated into the definitions of the string parameters which
appear in Eq. (\ref{stringLgaussian}).

\subsection{Renormalization and Phase Diagram}
\label{renormGauss}

In order to derive the renormalization equations for the electronic
parameters we integrate over the string configurations. Using a cumulant
expansion to average the electronic part of the action, Eq. (\ref{Lefluct}),
we find that to second order in $\lambda/a$, the result is of the ordinary
Luttinger type, Eq. (\ref{electLrigid}). In the averaged action the
velocities $v_\alpha$ are related to the values $v_{\alpha x}$ they
would have in a static straight string of length $L_x$ with Fermi
velocity $v_{Fx}$ and the same interactions, according to
\be
\label{v}
{v}_\alpha=v_{\alpha x}-\left(\frac{\lambda}{a}\right)^2\left[
\frac{v_{\alpha x}}{2}+\frac{v_{Fx}}{4}
\left(\frac{1}{K_{\alpha_x}}+K_{\alpha_x}\right)
\right] \; ,
\ee
while the Luttinger parameters and $g_{1\perp}$ are renormalized from
their values in the straight system as
\ba
\label{Kg}
\nonumber
K_\alpha&=&K_{\alpha x}-\left(\frac{\lambda}{a}\right)^2\frac{v_{Fx}}
{v_{\alpha x}}\left(1-K_{\alpha x}^2\right) \; ,\\
g_{1\perp}&=&g_{1\perp
x}-\frac{1}{2}\left(\frac{\lambda}{a}\right)^2 g_{1\perp x} \; . \ea
This renormalization is geometric in origin, as explained in
Ref.~\onlinecite{string}, and is absent (apart from a residual
correction to the velocities), when one keeps the electronic density
along the string, rather than the number of electrons, fixed. We
will therefore not discuss it further. Higher order terms in the
small parameter $\lambda/a$, which are generated in the averaging
process, are typically non-local and can be shown to be irrelevant
in the renormalization group sense.

Since in the present model $\langle[u(x,\tau)-u(x,\tau')]^2\rangle=
2\lambda^2\ln(1+v_u|\tau-\tau'|/a)$, the function
\be
\label{FGauss}
F(\tau)=\frac{1}{2\sqrt{\pi}\lambda}\ln^{-1/2}\left(1+\frac{v_u|\tau|}{a}
\right) \; ,
\ee
which appears in
the averaged disorder part of the action, Eqs. (\ref{barg}-\ref{effectiveS}),
lacks a cutoff independent length scale, nor is it a power law. Consequently,
an attempt to carry out a renormalization procedure for a $\delta$-correlated
Gaussian disorder, as was done for the rigid string, runs into the problem of
not yielding a simple multiplicative renormalization. Stated differently,
we are unable to assign a single scaling dimension to the disorder operator.
In order to overcome this difficulty we consider instead the more general
case of correlated impurity potentials
\ba
\label{corrdis}
\nonumber
\overline{\langle\eta(x,y)\eta(x',y')\rangle}&=&\delta(x-x')D_f(y-y') \\
\nonumber
&=&\delta(x-x')\int dk D_f(k)e^{ik(y-y')} \; , \\
\overline{\langle\xi^*(x,y)\xi(x',y')\rangle}&=&\delta(x-x')D_b(y-y') \\
\nonumber
&=&\delta(x-x')\int dk D_b(k)e^{ik(y-y')} \; ,
\ea
and employ functional renormalization to study the flow of $D_{b,f}(y-y')$.
As a result of the different disorder correlations one should replace
\ba
\label{replaceF}
\nonumber
D_{b,f}F(\tau)&\rightarrow&\int dk D_{b,f}(k)e^{-\frac{k^2}{2}
\langle[u(0,\tau)-u(0,0)]^2\rangle} \\
&=&\int dk D_{b,f}(k)\left(\frac{v_u|\tau|}{a}\right)^{-\lambda^2 k^2} \; ,
\ea
in Eqs. (\ref{effectiveS}-\ref{barvK}). Note that the last equality holds
approximately for $v_u|\tau|/a>1$, and is therefore appropriate for use in
the effective action, Eq. (\ref{effectiveS}). For a consistent treatment
we have to use the same form in the relations, Eq. (\ref{barvK}), where in
order to assure convergence of the $\tau$ integral we need to assume
$D_{b,f}(k)=0$ for $|k|>\lambda^{-1}$. Since at the bare level our stiff
string satisfies $\lambda\ll a$, and, as we shall demonstrate, $\lambda$
tends to decrease in the course of the renormalization, this requirement
is not stringent. The fact that the temporal correlations introduced by
each disorder Fourier component are algebraic makes it possible to follow
the same renormalization scheme we have used in the case of the rigid
string. Finally, we comment that because both the electronic and the
disorder parts of the action contain the string coordinates, averaging
over the latter introduces cross-terms between the two parts. These terms,
however, are higher derivative terms compared to those originating
from the averaged disorder part and are therefore less relevant. We will
therefore neglect them in the following.

Introducing the dimensionless disorder components
\ba
\label{dimdis}
\nonumber
{\cal D}_b(k)&=&\frac{2D_b(k)a}{\pi v_s^2 \lambda}\left(\frac{v_s}{v_c}
\right)^{K_c}\left(\frac{v_u}{v_s}\right)^{-\lambda^2 k^2} \; , \\
{\cal D}_f(k)&=&\frac{2D_f(k)a}{\pi v_c^2 \lambda}
\left(\frac{v_u}{v_s}\right)^{-\lambda^2 k^2} \; ,
\ea
we obtain the following renormalization equations
\begin{widetext}

\ba
\label{RGgauss}
\nonumber
\frac{dK_c}{d\ell}&=&-\frac{1}{2}\frac{v_c}{v_s}\int dk\lambda
\left[K_c^2-\frac{1}{2}\left(\frac{v_c}{v_s}\right)^{K_c-2}
\left(\frac{3-K_c-K_s-y-\lambda^2 k^2}{1-\lambda^2k^2}\right)(1+K_c^2)\right]
{\cal D}_b(k) \; ,\\
\nonumber
\frac{dv_c}{d\ell}&=&-\frac{1}{2}\frac{v_c^2}{v_s}\int dk\lambda
\left[K_c-\frac{1}{2}\left(\frac{v_c}{v_s}\right)^{K_c-2}
\left(\frac{3-K_c-K_s-y-\lambda^2 k^2}{1-\lambda^2k^2}\right)\left(
\frac{1}{K_c}-K_c\right)\right]
{\cal D}_b(k) \; ,\\
\nonumber
\frac{dK_s}{d\ell}&=&-\frac{1}{2}y^2 K_s^2
-\frac{1}{2}K_s^2\int dk\lambda
{\cal D}_b(k)+\frac{1}{4}\left(\frac{v_c}{v_s}\right)^{K_c}\int dk\lambda
\frac{(3-K_c-K_s-y-\lambda^2 k^2)(1+K_s^2)+4K_s^2 y}{1-\lambda^2k^2}
{\cal D}_b(k) \; ,\\
\frac{dv_s}{d\ell}&=&-\frac{v_s}{2}\int dk \lambda
\left[K_s-\frac{1}{2}\left(\frac{v_c}{v_s}\right)^{K_c}
\left(\frac{3-K_c-K_s-y-\lambda^2 k^2}{1-\lambda^2k^2}\right)
\left(\frac{1}{K_s}-K_s\right)\right]{\cal D}_b(k) \; ,\\
\nonumber
\frac{dy}{d\ell}&=&(2-2K_s)y-\left\{1-\left(\frac{v_c}{v_s}\right)^{K_c}
\left[1-\frac{y}{4}\left(\frac{1}{K_s}-K_s\right)\right]\right\}
\int dk\lambda {\cal D}_b(k) \\
\nonumber
&+&\frac{1}{4}\left(\frac{v_c}{v_s}\right)^{K_c}
\left\{4(K_s-K_c)-y\left[1-K_s^2+(2-K_c)\left(\frac{1}{K_s}-K_s\right)
\right]\right\}\int dk\lambda \frac{{\cal D}_b(k)}{1-\lambda^2 k^2} \; ,\\
\nonumber
\frac{d{\cal D}_f(k)}{d\ell}&=&(1-\lambda^2 k^2){\cal D}_f(k) \; , \\
\nonumber
\frac{d{\cal D}_b(k)}{d\ell}&=&(3-K_c-K_s-y-\lambda^2 k^2){\cal D}_b(k) \; .
\ea
The flow equations for the string parameters may be derived along the
lines indicated in Appendix \ref{stringRG} with the result
\ba
\label{RGstGauss}
\nonumber
&&\!\!\!\!\!\frac{d\lambda}{d\ell}=-\frac{1}{4}\frac{v_u}{v_s}\lambda
\left\{\left[1-f(v,K)\right]\int dk\lambda^3 k^2{\cal D}_b(k)
-f(v,K)(2-K_c-K_s-y)\int  dk\lambda \frac{k^2\lambda^2}{1-k^2\lambda^2}
\left[1-e^{-(1-k^2\lambda^2)\ell}\right]{\cal D}_b(k)\right\}  , \\
\nonumber
&&\!\!\!\!\!\frac{dv_u}{d\ell}=-\frac{1}{2}\frac{v_u^2}{v_s}
\left\{\left[1-f(v,K)\right]\int dk\lambda^3 k^2{\cal D}_b(k)
-f(v,K)(2-K_c-K_s-y)\int  dk\lambda \frac{k^2\lambda^2}{1-k^2\lambda^2}
\left[1-e^{-(1-k^2\lambda^2)\ell}\right]{\cal D}_b(k)\right\}  , \\
\ea

\end{widetext}
where $f(v,K)$ is given by Eq. (\ref{fvK}).

It is evident from the flow equation for the disorder that high-wavevector
components in the correlation function of the scattering potential are
less relevant. Consequently, the boundary separating the regions
of relevant and irrelevant disorder is determined by the lowest wavevector
components that are present in the correlation function. For a
very weak bare disorder which is $\delta$-correlated in real space,
this would mean that while it acquires, in the course of the renormalization,
correlations over longer length-scales, the critical point is still
given by the condition $3-K_c-K_s^*=0$, as in the cases of the static
and rigidly-oscillating strings. The $k$-dependence of the disorder scaling
dimension also implies that the flow of the electronic parameters in
Eq. (\ref{RGgauss}) is dominated by the $k\sim 0$ components, for
which it coincides with that of a static string. Therefore, the effects
of string fluctuations enters predominantly via the dependence of the
dimensionless disorder strength, Eq. (\ref{dimdis}), on $\lambda^{-1}$.
As fluctuations increase, and with them $\lambda$, the initial conditions
of the flow, for a given scattering potential, shift towards lower values
of ${\cal D}_b$. As a result the extent of the region in which the
disorder is irrelevant increases as well. Contrary to the case of the
rigid oscillating string there is no available solution which describes the
present system in the region of relevant disorder. We are therefore unable
to positively identify the nature of the strongly disordered phase. It
seems likely, however, that it is localized with magnetic properties
which depend on the flow of $y$ in the manner discussed at the end of the
previous section.

Considering the impurity forward scattering contribution to the suppression
of the CDW and SDW correlations we obtain, after taking
$D_f(k)=\frac{D_f}{2\pi}$, the following results for the function
$W(x,\tau)$ defined previously
\be
\label{WxGauss}
W(x,0)=\left\{
\begin{array}{l@{\quad  \quad}l}
\frac{3}{\sqrt{2\pi}}\frac{D_f K_c^2}{v_c v_u}\frac{a}{\lambda}
\sqrt{\frac{v_u x}{v_c a}}+{\cal O}(x^2) & \!\!\!\!\!\!\!\!\!
:|x|\ll \frac{v_c}{v_u}a \\
\frac{1}{\sqrt{\pi}}\frac{D_f K_c^2}{v_c^2}\frac{|x|}{\lambda} & \\
\times\!\left[\ln^{-\frac{1}{2}}\left|\frac{v_u x}{v_c a}\right|+{\cal O}
(\ln^{-\frac{3}{2}}|x|)\right] &\!\!\!\!\!\!\!\!\!:|x|\gg \frac{v_c}{v_u}a
\end{array}
\right. ,
\ee
and
\be
\label{WtGauss}
W(0,\tau)=\left\{
\begin{array}{l@{\quad  \quad}l}
\frac{1}{\sqrt{\pi}}\frac{D_f K_c^2}{v_c v_u}\frac{a}{\lambda}
\sqrt{\frac{v_u |\tau|}{a}}+{\cal O}(\tau^2) &
\!\!:v_u|\tau|\ll a \\
\frac{4.95}{4\pi^{\frac{3}{2}}}\frac{D_f K_c^2}{v_c \lambda}|\tau|
\ln^{-\frac{3}{2}}\left|\frac{v_u \tau}{a}\right| & \\
+{\cal O}\left(\ln^{-\frac{1}{2}}\left|\frac{v_u \tau}{a}\right|\right)
&\!\!:v_u|\tau|\gg a
\end{array}
\right. .
\ee

\section{Floppy String}
\label{sec:Floppy}

\subsection{The Model}
\label{modelFloppy}

An obvious way to increase the degree of string fluctuations in the
model is to replace the elastic energy term in the string Lagrangian,
Eq. (\ref{stringLgaussian}), by a new term involving higher spatial
derivatives of the displacement function $u$. Such a simple modification,
however, is insufficient, since, as we have demonstrated in the discussion
preceding Eq. (\ref{sigmarenorm}), the dependence of the electronic
kinetic energy on the string length will induce such an elastic term
even if it is absent from the bare string dynamics\cite{elasticom}.
In order to avoid this effect we need to consider a string of fixed
length $L$ and average electronic density $n$. We assume that the string
is held fixed at one of its end points (which we take as the origin),
and that its motion is constrained to be periodic in the $y$ direction,
i.e., that both ends are always at $y=0$. Since one end is free to slide
along the $x$ direction we are unable to characterize the configuration
of the string by a single displacement function $u(x,\tau)$. Instead we
use the arc-length  $s\in [0,L]$ to parameterize the string position
$\vec{R}(s,\tau)=[X(s,\tau),Y(s,\tau)]$ in the plane. Taking the bending
energy of the string to depend on its curvature
$\sqrt{\left(\partial^2 X/\partial s^2\right)^2+
\left(\partial^2 Y/\partial s^2\right)^2}$ one obtains
\be
\label{stringLfloppy}
L_s=\int_0^{L} ds \left[\frac{\rho}{2}\left(\frac{\partial \vec{R}}
{\partial \tau}\right)^2+\frac{\gamma}{2}
\left(\frac{\partial^2 \vec{R}}{\partial s^2}\right)^2\right]  \; ,
\ee
together with the constraint
\be
\label{constraint}
(\partial_s X)^2+(\partial_s Y)^2=1 \; ,
\ee
reflecting our choice of parameterization.

To make progress we need to assume that the string is massive and possesses
a large bending modulus $\gamma$, such that its fluctuations about the $x$
axis take place over large length scales. Specifically we require that
\be
\label{barlambda}
\bar{\lambda}\equiv\frac{\lambda}{a}=\frac{1}{a\sqrt{\rho\gamma}}\ll 1 \; .
\ee
As we now demonstrate, under this condition the constraint, Eq.
(\ref{constraint}), may be solved approximately
$\partial_s X\simeq 1-\frac{1}{2}(\partial_s Y)^2$, to imply,
given the specified boundary conditions, that
\ba
\label{DeltaX}
\nonumber
X(s,\tau)&=&s+\Delta X(s,\tau) \\
&=&s-\frac{1}{2}\int_0^s ds' [\partial_{s'}Y(s',\tau)]^2 \; .
\ea
Using this result in the string Lagrangian, Eq. (\ref{stringLfloppy}), one
indeed finds\cite{schemecom}, to lowest order in $\bar{\lambda}$,
\be
\label{averageslope}
\langle(\partial_s Y)^2\rangle = \sqrt{\frac{\bar{\lambda}}{2\pi^2}} \; ,
\ee
in agreement with the assumption which led to the expansion, Eq.~(\ref{DeltaX}).

As long as the system maintains, in addition to condition
(\ref{barlambda}), that its typical string wave velocity (given in
this model by $\gamma$) is such that $\gamma/a<v_F k_F$, we may
neglect backward scattering by string fluctuations  and describe the
low-energy electronic dynamics by the Lagrangian
\begin{widetext}
\ba
\label{Lefloppy}
\nonumber
L_e=\int_0^{L} ds \Bigg\{&&
\sum_{\alpha=c,s}\left[-i\partial_\tau\phi_\alpha\partial_s\theta_\alpha
+\frac{v_\alpha K_\alpha}{2}(\partial_s\theta_\alpha)^2+\frac{v_\alpha}
{2K_\alpha}(\partial_s\phi_\alpha)^2\right]
+\frac{g_{1\perp}}{2(\pi a)^2}\cos(\sqrt{8\pi}\phi_s) \\
&&-e{\cal A}_0\left(n + \sqrt{\frac{2}{\pi}} \partial_s\phi_c \right)
-v_c K_c\sqrt{\frac{2}{\pi}}\frac{e}{c}{\cal A}_1
\partial_s\theta_c +v_c K_c\frac{e^2}{\pi c^2}{\cal A}_1^2\Bigg\} \, ,
\ea

\end{widetext}
where, provided the electrons interact through contact interactions
$V(s-s')\propto\delta(s-s')$, the velocities and Luttinger
parameters are given by Eq. (\ref{vK}). While in the arc-length
parameterization the metric remains trivial, the electrons are still
coupled to gauge potentials, which to second order in derivatives of
$Y$ read 
\ba 
\label{approxpot} 
\nonumber
{\cal A}_0(s,\tau)&=&\!-\frac{m}{2e}\left(\partial_\tau\vec{R}\right)^2
\simeq\frac{m}{2e}\left(\partial_\tau Y\right)^2  , \\
{\cal A}_1(s,\tau)&=&\!-\frac{imc}{e}\partial_s\vec{R}
\cdot\partial_\tau\vec{R} \\
\nonumber
&\simeq& -\frac{imc}{e}\left(
\partial_s Y\partial_\tau Y+\partial_\tau \Delta X \right)  .
\ea
Note that the term proportional to $n$ in Eq. (\ref{Lefloppy}) leads to
the renormalization of the string mass density according to
$\rho\rightarrow\rho+mn$. However, as expected, the coupling between the
string and the electrons does not modify the form of the string elastic
energy term.

For a $\delta$-correlated random potential the averaged disorder part
of the action contains the factor
\ba
\label{disorderav}
\nonumber
&&\!\!\!\!\!
D_{f,b}\delta[X(s,\tau)-X(s',\tau')]\delta[Y(s,\tau)-Y(s',\tau')] \simeq \\
\nonumber
&&\!\!\!\!\!
D_{f,b}\delta(s-s')\delta[Y(s,\tau)-Y(s,\tau')]\left\{1+\frac{1}{2}
\left[\partial_s Y(s,\tau)\right]^2\right\} , \\ 
\ea
in which, to lowest order in $D_{f,b}\bar{\lambda}$, we may replace the curly
brackets by 1.

\subsection{Renormalization and Phase Diagram}
\label{renormFloppy}

Integrating out the string degrees of freedom from the electronic
Lagrangian, Eq. (\ref{Lefloppy}), yields at the level of the first
cumulant (quadratic in derivatives of $Y$), a simple Luttinger liquid
action, Eq. (\ref{electLrigid}). The contribution from the second
cumulant (fourth order in derivatives of $Y$) modifies the values of
$v_c$, $K_c$, and the coefficient of the simplectic term in
the action by a small amount of order $(mv_c\lambda)^2\bar{\lambda}^{1/2}
(\gamma/v_c)K_c$. In addition, however, it also introduces the terms
\ba
\label{relterms}
\nonumber
&&\frac{\beta L}{8\pi}\frac{(mv_c\lambda K_c)^2}{\sqrt{\lambda\gamma}}
{\Bigg \{} \frac{1}{2}
\sum_{k,k'\neq0,\omega}|\omega|^{3/2}
\theta_c(k,\omega)\theta_c(k',-\omega) \\
\nonumber
&&+\sum_{k\neq0,\omega}\frac{\omega^2}
{\left[\lambda^2\gamma^2 k^4+\omega^2\right]^{1/4}}
\cos\left[\frac{1}{2}\arctan\left(\frac{2|\omega|}
{\lambda\gamma k^2}\right)\right] \\
&&\hspace{1cm}\times\theta_c(k,\omega)\theta_c(-k,-\omega){\Bigg \}} \; ,
\ea
of which the first originates from the breaking of translation invariance
along the $x$ direction due to the boundary conditions on the string.
These terms are relevant in the long-wavelength limit and dominate
over the quadratic terms in the Luttinger action for $k\rightarrow 0$ and
$\omega<\left(\frac{K_c}{4\pi}\right)^2(mv_c\lambda)^4\frac{v_c^2}
{\lambda\gamma}\equiv v_s/\bar{a}$, with the effect of suppressing
slow fluctuations of $\theta_c$ (effectively increasing the value of $K_c$
at long time scales). Based on the flow equations we have derived before
we expect that this fact would lead to a reduction in the scaling
dimension of the backward scattering disorder at scales larger than
$\bar{a}$. In the following we will ignore the contribution, Eq.
(\ref{relterms}), (and possibly other relevant terms which emerge at
higher orders in the cumulant expansion), thereby limiting
our renormalization treatment to the range $a<\bar{a}$. In that sense
our results should be viewed as an upper limit on the domain of relevant
disorder.

Finally, averaging the disorder part of the action over the string dynamics
one obtains an expression similar to that of Eq. (\ref{effectiveS}) with
\be
\label{Ffloppy}
F(\tau)=\frac{1}{\lambda}\left(\frac{\lambda}{4\pi\gamma|\tau|}\right)
^{1/4}  ,
\ee
which can be treated according to the lines indicated in the appendices
to yield the following renormalization equations for the electronic and
string parameters
\begin{widetext}

\ba
\label{RGfloppy}
\nonumber
\frac{dK_c}{d\ell}&=&-\frac{1}{2}\frac{v_c}{v_s}
\left[K_c^2-\frac{2}{3}\left(\frac{v_c}{v_s}\right)^{K_c-2}
\left(\frac{11}{4}-K_c-K_s-y\right)(1+K_c^2)\right]{\cal D}_b \; ,\\
\nonumber
\frac{dv_c}{d\ell}&=&-\frac{1}{2}\frac{v_c^2}{v_s}
\left[K_c-\frac{2}{3}\left(\frac{v_c}{v_s}\right)^{K_c-2}
\left(\frac{11}{4}-K_c-K_s-y\right)\left(\frac{1}{K_c}-K_c\right)
\right]{\cal D}_b \; ,\\
\nonumber
\frac{dK_s}{d\ell}&=&-\frac{1}{2}y^2 K_s^2
+\frac{1}{3}\left(\frac{v_c}{v_s}\right)^{K_c}
\left[\left(\frac{11}{4}-K_c-K_s-y\right)(1+K_s^2)
-\frac{3}{2}\left(\frac{v_s}{v_c}\right)^{K_c}K_s^2+4K_s^2 y\right]
{\cal D}_b \; , \\
\frac{dv_s}{d\ell}&=&-\frac{1}{2}
\left[K_s-\frac{2}{3}\left(\frac{v_c}{v_s}\right)^{K_c}
\left(\frac{11}{4}-K_c-K_s-y\right)\left(\frac{1}{K_s}-K_s\right)
\right]v_s{\cal D}_b \; ,\\
\nonumber
\frac{dy}{d\ell}&=&(2-2K_s)y+\frac{1}{3}\left(\frac{v_c}{v_s}\right)^{K_c}
\!\left\{3\left[1-\left(\frac{v_s}{v_c}\right)^2\right]
+4(K_s-K_c)-y\left[1-K_s^2+\left(\frac{11}{4}-K_c\right)
\left(\frac{1}{K_s}-K_s\right)\right]\right\}{\cal D}_b \; , \\
\nonumber
\frac{d{\cal D}_f}{d\ell}&=&\frac{3}{4}{\cal D}_f \; , \\
\nonumber
\frac{d{\cal D}_b}{d\ell}&=&\left(\frac{11}{4}-K_c-K_s-y\right)
{\cal D}_b \; . \\
\nonumber
\frac{d\bar{\lambda}}{d\ell}&=&-\bar{\lambda}
-\left(\frac{\gamma}{\pi v_s}\right)^{1/2}
\left\{ \frac{1}{4}-f(v,K)\left[\frac{1}{4}+(2-K_c-K_s-y)\left(1-e^{-\ell/4}
\right)\right]\right\}\bar{\lambda}^{3/2}{\cal D}_b \; , \\
\nonumber
\frac{d\gamma}{d\ell}&=&-\frac{1}{2}
\left[K_s-\frac{2}{3}\left(\frac{v_c}{v_s}\right)^{K_c}
\left(\frac{11}{4}-K_c-K_s-y\right)\left(\frac{1}{K_s}-K_s\right)
\right]\gamma{\cal D}_b \; ,
\ea

\end{widetext}
where $f(v,K)$ is given in Eq. (\ref{fvK}), and where the
dimensionless disorder amplitudes for this model are defined
according to \ba \label{dimDfloppy} \nonumber {\cal
D}_b&=&\frac{\sqrt{2}D_b}{\pi^{5/4}v_s^2}\left(\frac{a}{\lambda}\right)
^{3/4}\left(\frac{v_s}{\gamma}\right)^{1/4}\left(\frac{v_s}{v_c}\right)^{K_c}
\; , \\
{\cal D}_f&=&\frac{\sqrt{2}D_f}{\pi^{5/4}v_c^2}\left(\frac{a}{\lambda}\right)
^{3/4}\left(\frac{v_s}{\gamma}\right)^{1/4} \; .
\ea

The enhanced fluctuations of the floppy string do result in a shift of the
critical point which demarcates the border between regions of relevant and
irrelevant flow for weak initial disorder potential. The position of this
point is given by the condition $11/4-K_c-K_s^*=0$, which for a
spin-rotationally invariant system and in the case where $y$ flows to zero
implies that the system is delocalized for $K_c>7/4$ (we note once again
that due to our inability to treat terms such as Eq. (\ref{relterms}) this
should be viewed as a lower limit on the extent of the delocalized region.)
This is to be contrasted with the previously studied models for which
delocalization occurs under similar conditions for $K_c>2$. For a generalized
model of a stiff and massive fixed-length string with elastic energy that is
proportional to $(\partial^n{\vec R}/\partial s^n)^2$ one finds (ignoring
potentially relevant terms which may arise in the course of the
renormalization) a delocalized phase for $K_c>(3n+1)/2n$. In close
analogy to the situation in the other models we have considered in
this study the fact that
${\cal D}_b\propto \lambda^{-3/4}\gamma^{-1/4}=\rho^{3/8}\gamma^{1/8}$
implies that enhancing the string fluctuations by making $\rho$ and $\gamma$
smaller has the effect of shifting the initial conditions of the flow
towards smaller values of ${\cal D}_b$, thereby increasing the extent of
the region in parameter space in which the disorder is irrelevant.

As far as the CDW and SDW correlations are concerned, we find that their decay
due to forward-scattering by disorder is given by
\ba
\label{Wfloppy}
W(x,0)&=&\frac{7\pi}{6\cos\left(\frac{\pi}{8}\right)}K_c^2{\cal D}_f
\left(\frac{v_c}{v_s}\right)^{1/4}\left(\frac{|x|}{a}\right)^{3/4} \; , \\
\nonumber
W(0,\tau)&=&\frac{\sqrt{2}\pi}{6\cos^2\left(\frac{\pi}{8}\right)}
K_c^2{\cal D}_f
\left(\frac{v_c}{v_s}\right)^{1/4}\left(\frac{v_c|\tau|}{a}\right)^{3/4} \; ,
\ea
at least for $|x|,v_s|\tau|<\bar{a}$ such that we may neglect the higher
order corrections to the electronic effective action, Eq. (\ref{relterms}).

\section{Conclusion and Discussion}

In this paper we have set out to explore the way in which geometrical
fluctuations of a one-dimensional interacting system affect its
renormalization flow in the presence of a random scattering potential.
We have found that by inducing temporal variations in the disorder, as
seen in the reference frame of the electrons, the geometrical fluctuations
increase the region in parameter space where the disorder is an irrelevant
perturbation. This is a result of processes in which potential wells
diminish in time thereby releasing electrons that were trapped inside them.
Other effects, such as decoherence of interference patterns which lead to
localization also exist but are not captured by our treatment. We expect,
however, that the latter are less important in one-dimension.
Notwithstanding, unless the fluctuations are strong enough,
the general features of the phase diagram are unchanged and the
localization-delocalization transition for weak bare disorder remains at
$K_c=2$, in the case of a gapless spin-invariant system. In the hierarchy
of models studied by us only in the floppy string model the critical
point has shifted towards weaker values of attractive interactions.

Our results thus demonstrate the relative robustness of
disorder-induced localization-effects in strictly one-dimensional
systems. In order to suppress such localization effects it thus
seems necessary to introduce some degree of two dimensionality into 
the system. This can be achieved by allowing the electrons to hop
from one one-dimensional chain to the other. A simple realization is
provided by ladders. For example, in the maximally gapped phase of a
two-leg ladder\cite{2legdis} small disorder is an irrelevant
perturbation when the Luttinger parameter, $K_{c+}$, associated with
the gapless total charge mode, is greater than 3/2, thus smaller
than for a single chain. In a two leg bosonic ladder \cite{2legbos}
or in the spin-gapped phase of a four-leg ladder\cite{optimal}
disorder is irrelevant for $K_{c+}>3/4$. The results of introducing
geometrical fluctuations to such ladder systems are yet to be
studied. Understanding such effects would be very interesting,
since, despite the fact that the disorder is an irrelevant
perturbation, a finite amount of disorder is still quite efficient
in killing the superconducting phase\cite{2legdis}. As we have seen
for the case of a single chain, string fluctuations are quite
efficient in diminishing the size of the localized region even
when it is not able to change the critical point. One could thus
expect particularly interesting effects of such fluctuations in the
case of ladders.

\acknowledgments
This research was supported by the Israel Science Foundation
(grant No. 193/02-1). Part of this work was supported by the Swiss
National Science Foundation under Division II.

\appendix
\section{The impurity forward scattering contribution to the flow equations}
\label{forwardRG}

Here we briefly demonstrate how the forward scattering component of the
disorder influence the renormalization of the electronic parameters of the
rigid string effective action, Eq. (\ref{effectiveS}). To this end, we
consider the correlation function $R_c(x_1-x_2,\tau_1-\tau_2)=\overline{
\langle T_\tau e^{i\sqrt{2\pi}\phi_c(x_1,\tau_1)} e^{-i\sqrt{2\pi}
\phi_c(x_2,\tau_2)} \rangle}$. To first order in the disorder strength
one finds
\begin{widetext}

\ba
\label{fscontrib}
\nonumber
R_c(x_1-x_2,\tau_1-\tau_2)&=&e^{-F_c(\vec{r}_1-\vec{r}_2)}
\Bigg\{ 1-2D_f\left(\frac{\overline{K}_c}{2\pi}
\right)^2\sqrt{\frac{M\omega_0}{2\pi}}
\int dx\int _{|\tau-\tau'|>a/v_s} d\tau d\tau'
\frac{1}{\sqrt{1-e^{-\omega_0|\tau-\tau'|}}} \\
\nonumber
&\times&\left[\frac{x-x_1}{(x-x_1)^2+\overline{v}_c^2(\tau-\tau_1)^2}-
\frac{x-x_2}{(x-x_2)^2+\overline{v}_c^2(\tau-\tau_2)^2}\right] \\
&\times&\left[\frac{x-x_1}{(x-x_1)^2+\overline{v}_c^2(\tau'-\tau_1)^2}-
\frac{x-x_2}{(x-x_2)^2+\overline{v}_c^2(\tau'-\tau_2)^2}\right]
+\begin{array}{c}{\rm backward\;\, scattering}\\{\rm  contribution}
\end{array}\Bigg\} \; ,
\ea
with
\ba
\label{Fc}
F_c(\vec{r})&=&\frac{\overline{K}_c}{2}\ln\left(\frac{x^2+
\overline{v}_c^2\tau^2}{a^2}\right)+\overline{f}_c\frac{
\overline{v}_c^2\tau^2}{x^2+\overline{v}_c^2\tau^2} \; ,
\ea
where initially $\overline{f}_c=0$. After carrying out the integration
over $x$ and the center of mass coordinate $(\tau+\tau')/2$ in Eq.
(\ref{fscontrib}), one obtains
\ba
\label{fscontrib2}
\nonumber
R_c(x_1-x_2,\tau_1-\tau_2)&=&e^{-F_c(\vec{r}_1-\vec{r}_2)}
\Bigg\{ 1-\frac{\overline{v}_c\overline{K}_c^2{\cal D}_f}{4}
\int _{|\eta|>a/v_s}\!\frac{d\eta}{a}
\frac{1}{\sqrt{1-e^{-\frac{\varpi v_s}{a}|\eta|}}} \Bigg[
\ln\left[\frac{(x_1-x_2)^2+\overline{v}_c^2
(\eta+\tau_1-\tau_2)^2}{\overline{v}_c^2\eta^2}\right]  \\
&+&\frac{2(x_1-x_2)^2}{(x_1-x_2)^2
+\overline{v}_c^2(\eta+\tau_1-\tau_2)^2} \Bigg]
+\begin{array}{c}{\rm backward\;\, scattering}\\{\rm  contribution}
\end{array}\Bigg\} \; .
\ea

\end{widetext}
The flow equations are derived by requiring that the long range
behavior of $R_c$ remains invariant under a change in the cutoff
$a=\rightarrow a+da=a(1+d\ell)$. In order to maintain such
invariance the missing contribution from the integration over
$(a+da)/v_s>|\eta|>a/v_s$ should be compensated by an appropriate
change in $\overline{K}_c$ and $\overline{f}_c$ (the latter is
related\cite{gs} to a change in $\overline{v}_c$). The invariance of
the remaining integral over $|\eta|>(a+da)/v_s$ determines the flow
of ${\cal D}_f$ and $\varpi$. The equations for $\overline{K}_s$,
$\overline{v}_s$ and $\overline{y}$ can be derived in a similar
manner by considering the correlation function
$R_s(x_1-x_2,\tau_1-\tau_2)=\overline{\langle T_\tau e^{i\sqrt{2\pi}
\phi_s(x_1,\tau_1)} e^{-i\sqrt{2\pi}\phi_s(x_2,\tau_2)} \rangle}$.
They, however, do not contain the forward scattering amplitude.
Including the contribution stemming from the backward scattering
piece, as described in Ref. \onlinecite{gs}, one arrives at 
\ba
\label{RGbar} \nonumber
\frac{d\overline{K}_c}{d\ell}&=&-\frac{v_c}{v_s}\frac{1}{\sqrt{1-e^{-\varpi}}}
\left(\frac{{\cal D}_b}{2}-{\cal D}_f\right)K_c^2 \; , \\
\nonumber
\frac{d\overline{v}_c}{d\ell}&=&-\frac{v_c^2}{v_s}\frac{1}
{\sqrt{1-e^{-\varpi}}}\left(\frac{{\cal D}_b}{2}+{\cal D}_f\right)K_c \; , \\
\nonumber
\frac{d\overline{K}_s}{d\ell}&=&-\frac{1}{2}\overline{y}^2\overline{K}_s^2
-\frac{{\cal D}_b K_s^2}{2\sqrt{1-e^{-\varpi}}} \; , \\
\nonumber
\frac{d\overline{v}_s}{d\ell}&=&-\frac{{\cal D}_b K_s v_s}
{2\sqrt{1-e^{-\varpi}}} \; , \\
\frac{d\overline{y}}{d\ell}&=&(2-2\overline{K}_s)\overline{y}-
\frac{{\cal D}_b}{\sqrt{1-e^{-\varpi}}} \; , \\
\nonumber
\frac{d{\cal D}_b}{d\ell}&=&(3-K_c-K_s-y){\cal D}_b \; , \\ 
\nonumber
\frac{d{\cal D}_f}{d\ell}&=&{\cal D}_f \; , \\ 
\nonumber
\frac{d\varpi}{d\ell}&=&\varpi+{\cal O}({\cal D}) \; .
\ea
These results are readily extended to the cases of the elastic and floppy
strings by replacing the expression for $F(\tau)$ in Eqs. (\ref{fscontrib},
\ref{fscontrib2}) with its appropriate form for these models, Eqs.
(\ref{replaceF},\ref{Ffloppy}).

\appendix
\setcounter{section}{1}
\section{The flow equations for the string parameters}
\label{stringRG}

To obtain the flow equation for the rigid string parameters we consider the
correlation function $R_u(\tau_1-\tau_2)=\overline{\langle T_\tau u(\tau_1)
u(\tau_2)\rangle}$, where the average is with respect to the effective
action, Eq. (\ref{effSs}). To first order in the disorder strength one finds
\begin{widetext}

\ba
\label{Ru}
\nonumber
R_u(\Delta r)&=&\lambda^2 e^{-\varpi\frac{|\Delta r|}{a}}\Bigg\{
1-{\cal D}_b{\cal L}f(v,K)\int_{a_0}^a\frac{d\eta}{a}\frac{1}{\left(
1-e^{-\frac{\varpi\eta}{a}}\right)^{3/2}}\left(\frac{a}{\eta}\right)^2
\left[\left(\frac{1}{\varpi}+\frac{|\Delta r|}{a}\right)\left(1-\cosh
\frac{\varpi\eta}{a}\right)+\frac{\eta}{a}\sinh\frac{\varpi\eta}{a}\right] \\
\nonumber
&-&{\cal D}_b{\cal L}\int_{a}^{|\Delta r|}\frac{d\eta}{a}\frac{1}{\left(
1-e^{-\frac{\varpi\eta}{a}}\right)^{3/2}}\left(\frac{a}{\eta}\right)^{K_c+K_s}
\left[\left(\frac{1}{\varpi}+\frac{|\Delta r|}{a}\right)\left(1-\cosh
\frac{\varpi\eta}{a}\right)+\frac{\eta}{a}\sinh\frac{\varpi\eta}{a}\right] \\
&-&{\cal D}_f{\cal L}K_c\int_{a_0}^{|\Delta r|}\frac{d\eta}{a}\frac{1}{\left(
1-e^{-\frac{\varpi\eta}{a}}\right)^{3/2}}\left(\frac{a}{\eta}\right)^2
\left[\left(\frac{1}{\varpi}+\frac{|\Delta r|}{a}\right)\left(1-\cosh
\frac{\varpi\eta}{a}\right)+\frac{\eta}{a}\sinh\frac{\varpi\eta}{a}\right]
\Bigg \} \\
\nonumber
&-&\lambda^2{\cal L}\int_{|\Delta r|}^\infty\frac{d\eta}{a}\frac{1}{\left(
1-e^{-\frac{\varpi\eta}{a}}\right)^{3/2}}\left[{\cal D}_b\left(\frac{a}{\eta}
\right)^{K_c+K_s}+{\cal D}_f K_c\left(\frac{a}{\eta}\right)^2\right] \\
\nonumber
&\times&\left\{\left(\frac{1}{\varpi}+\frac{|\Delta r|}{a}\right)
e^{-\frac{\varpi|\Delta r|}{a}}
+ e^{-\frac{\varpi\eta}{a}}
\left[\frac{|\Delta r|}{a}\sinh\frac{\varpi|\Delta r|}{a}
-\left(\frac{1}{\varpi}+\frac{\eta}{a}\right)\cosh\frac{\varpi|\Delta r|}{a}
\right]\right\} \; ,
\ea

\end{widetext}
where $\Delta r \equiv v_s(\tau_1-\tau_2)$ is assumed larger than
$a$ and $f(v,k)$ is given in Eq. (\ref{fvK}).

The last two integrals in Eq. (\ref{Ru}) are invariant under a change in the
cutoff $a\rightarrow a+da$ provided one uses the previously derived
renormalization equations for the electronic and disorder parameters,
Eq. (\ref{RGrigid}), and assumes $d\varpi/d\ell=\varpi+{\cal O}({\cal D})$.
The remaining terms may be brought to the form $\lambda_{\rm eff}^2
e^{-\varpi_{\rm eff}\frac{|\Delta r|}{a}}$ with
\ba
\label{wleff}
\nonumber
\varpi_{\rm eff}&=&\varpi+{\cal D}_b{\cal L}\left[f(v,K)\int_{a_0}^a
\frac{d\eta}{a}\frac{1-\cosh\frac{\varpi\eta}{a}}
{\left(1-e^{-\frac{\varpi\eta}{a}}\right)^{3/2}}
\left(\frac{a}{\eta}\right)^2 \right. \\
&+&\left.\int_{a}^{|\Delta r|}\frac{d\eta}{a}\frac{1-\cosh\frac{\varpi\eta}{a}}
{\left(1-e^{-\frac{\varpi\eta}{a}}\right)^{3/2}}
\left(\frac{a}{\eta}\right)^{K_c+K_s} \right]  , 
\ea
\ba
\nonumber
\lambda_{\rm eff}^2&=&\lambda^2-\frac{{\cal D}_b{\cal L}\lambda^2}{\varpi} \\
\nonumber
&\times&\!\!\!\left[f(v,K)\int_{a_0}^a
\frac{d\eta}{a}\frac{1-\cosh\frac{\varpi\eta}{a}+\frac{\varpi\eta}{a}\sinh
\frac{\varpi\eta}{a}}
{\left(1-e^{-\frac{\varpi\eta}{a}}\right)^{3/2}}
\left(\frac{a}{\eta}\right)^2 \right. \\
\nonumber
\!\!&+&\left.\!\!\!\int_{a}^{|\Delta r|}\frac{d\eta}{a}
\frac{1-\cosh\frac{\varpi\eta}{a}
+\frac{\varpi\eta}{a}\sinh\frac{\varpi\eta}{a}}
{\left(1-e^{-\frac{\varpi\eta}{a}}\right)^{3/2}}\left(\frac{a}{\eta}\right)
^{K_c+K_s} \right] \!.
\ea
The renormalization equations for the string parameters,
Eq. (\ref{RGstrigid}), then follow from the requirement that
$R_u$ remains invariant under a change of the cutoff, i.e.,
$d\left(\frac{\varpi_{\rm eff}}{a}\right)/d\ell=0$ and
$d\lambda_{\rm eff}/d\ell=0$.

In the case of the Gaussian string we find
\begin{widetext}

\ba
\label{RuGauss}
\nonumber
\langle T_\tau u(x_1,\tau_1) u(x_2,\tau_2)\rangle&=& F_u(\vec{r}_1-\vec{r}_2)
\left\{1-\frac{v_u}{2v_s}\int dk \lambda^3k^2 K_c{\cal D}_f(k)
\int_{a_0}^\infty\frac{d\eta}{a}\left(\frac{\eta}{a}\right)^{-\lambda^2 k^2}
\right. \\
&-&\left.\frac{v_u}{2v_s}\int dk \lambda^3 k^2{\cal D}_b(k)\left[f(v,K)
\int_{a_0}^a \frac{d\eta}{a}\left(\frac{\eta}{a}\right)^{-\lambda^2 k^2} +
\int_a^\infty \frac{d\eta}{a}\left(\frac{\eta}{a}\right)^
{(2-K_c-K_s-\lambda^2 k^2)}\right]\right\} \; ,
\ea
where $f(v,K)$ is given by Eq. (\ref{fvK}), and
\be
\label{FuGauss}
F_u(\vec{r}_1-\vec{r}_2)=-\lambda^2\ln\left(\frac{2\pi}{L_x}\sqrt{
(x_1-x_2)^2+v_u^2(\tau_1-\tau_2)^2}\right)-
f_u\lambda^2\frac{v_u^2(\tau_1-\tau_2)^2}
{(x_1-x_2)^2+v_u^2(\tau_1-\tau_2)^2} \; ,
\ee

\end{widetext}
with the bare value $f_u=0$. The term containing ${\cal D}_f(k)$ in
Eq. (\ref{RuGauss}) is invariant under $a\rightarrow a+da$ provided
we use the flow equation for ${\cal D}_f(k)$, as given in Eq.
(\ref{RGgauss}), and assume $d\lambda/d\ell={\cal O}({\cal D})$,
$dv_u/d\ell={\cal O} ({\cal D})$, which is consistent with the final
result. The renormalization equation for the length $\lambda$, is
derived from the invariance requirement of the remaining logarithmic
terms in Eq. (\ref{RuGauss}) and using the flow equation for ${\cal
D}_b(k)$. In order to obtain the renormalization of the string wave
velocity we first note that $d(f_u\lambda^2) /d\ell=d\lambda^2/d\ell$.
Secondly, a renormalization $v_u\rightarrow v_u+ dv_u$ in the 
logarithmic part of $F_u$ is equivalent to a renormalization of
$f_u\lambda^2$ according to
$d(f_u\lambda^2)/d\ell=(\lambda^2/v_u)dv_u /d\ell$. Combining the
two we find that to first order in ${\cal D}_b$ \be
\label{renormrel}
\frac{dv_u}{d\ell}=\frac{2v_u}{\lambda}\frac{d\lambda}{d\ell} \; ,
\ee which results in the renormalization equations, Eq.
(\ref{RGstGauss}).

For the floppy string one obtains
\begin{widetext}
\ba
\label{RuFloppy}
&&\langle T_\tau u(s_1,\tau_1) u(s_2,\tau_2)\rangle= \frac{\bar{\lambda}a}
{4\pi}\int dq\frac{1}{q^2}e^{iq\Delta s-\gamma\bar{\lambda}a|\Delta\tau|
q^2}\\
\nonumber
&&\times {\Bigg \{} 1 -{\cal D}_b f(v,K)\left(\frac{v_s}
{4\pi\gamma\bar{\lambda}}\right)^{1/2}\int_{a_0}^a \frac{d\eta}{a}
\left(\frac{a}{\eta}\right)^{11/4}\left[\left(\frac{v_s}{\gamma\bar{\lambda}
q^4a^4}+\frac{v_s|\Delta\tau|}{q^2a^3}\right)\left(
1-\cosh\frac{\gamma\bar{\lambda}q^2a\eta}{v_s}\right)
+\frac{\eta}{q^2a^3}\sinh\frac{\gamma\bar{\lambda}q^2a\eta}{v_s}\right]\\
\nonumber
&&-{\cal D}_b\left(\frac{v_s}{4\pi\gamma\bar{\lambda}}\right)^{1/2}
\int_a^{v_s|\Delta\tau|} \frac{d\eta}{a}\left(\frac{a}{\eta}\right)^
{3/4+K_c+K_s}\left[\left(\frac{v_s}{\gamma\bar{\lambda}
q^4a^4}+\frac{v_s|\Delta\tau|}{q^2a^3}\right)\left(
1-\cosh\frac{\gamma\bar{\lambda}q^2a\eta}{v_s}\right)
+\frac{\eta}{q^2a^3}\sinh\frac{\gamma\bar{\lambda}q^2a\eta}{v_s}\right]\\
\nonumber
&&-{\cal D}_fK_c\left(\frac{v_s}{4\pi\gamma\bar{\lambda}}\right)^{1/2}
\int_{a_0}^{v_s|\tau_1-\tau_2|} \frac{d\eta}{a}\left(\frac{a}{\eta}\right)^
{11/4}\left[\left(\frac{v_s}{\gamma\bar{\lambda}
q^4a^4}+\frac{v_s|\Delta\tau|}{q^2a^3}\right)\left(
1-\cosh\frac{\gamma\bar{\lambda}q^2a\eta}{v_s}\right)
+\frac{\eta}{q^2a^3}\sinh\frac{\gamma\bar{\lambda}q^2a\eta}{v_s}\right]
{\Bigg \}}\\
\nonumber
&&-\left(\frac{v_s\bar{\lambda}}{64\pi^3\gamma}\right)^{1/2}
\int_{v_s|\Delta\tau|}^{\infty}\frac{d\eta}{a}\left[{\cal D}_b
\left(\frac{a}{\eta}\right)^{3/4+K_c+K_s}+{\cal D}_f K_c\left(\frac{a}{\eta}
\right)^{11/4}\right]\int dq \frac{1}{q^2}e^{iq\Delta s}\\
\nonumber
&&\times\left\{ e^{-\gamma\bar{\lambda}a|\Delta\tau|q^2}
\left(\frac{v_s}{\gamma\bar{\lambda}q^4a^4}+\frac{v_s|\Delta\tau|}{q^2a^3}
\right)-e^{\frac{\gamma}{v_s}\bar{\lambda}q^2a\eta}
\left[\frac{v_s|\Delta\tau|}{q^2a^3}\sinh\gamma\bar{\lambda}a
|\Delta\tau|q^2-\left(\frac{v_s}{\gamma\bar{\lambda}q^4a^4}+\frac{\eta}
{q^2a^3}\right)\cosh\gamma\bar{\lambda}a|\Delta\tau|q^2 \right]\right\} \; ,
\ea

\end{widetext}
where $\Delta s\equiv s_1-s_2$ and $\Delta\tau\equiv\tau_1-\tau_2$.
The $q$ integrals in the above are dominated by the small $q$
region, therefore allowing us to expand the hyperbolic functions in
the integrands. Using the flow equations for the electronic and
disorder parameters, Eqs. (\ref{RGfloppy}), and assuming
$d\bar{\lambda}/d\ell= -\bar{\lambda}+{\cal O}({\cal D})$ and
$d(\gamma/v_s)/d\ell=0$, the scale invariance of the last two
$\eta$ integrals in Eq. (\ref{RuFloppy}) is then readily verified.
The remaining two terms may be expressed as
$\frac{\bar{\lambda}_{\rm eff} a}{4\pi}\int
dq\frac{1}{q^2}e^{iq\Delta s -\gamma\bar{\lambda}_{\rm
eff}a\Delta\tau q^2}$ where \ba \label{efflamflop} \nonumber
\bar{\lambda}_{\rm
eff}=\bar{\lambda}&-&\frac{\sqrt{\gamma}\bar{\lambda}^{3/2} {\cal
D}_b}{4\sqrt{\pi v_s}} \left[f(v,K)\int_{a_0}^a
\frac{d\eta}{a}\left(
\frac{a}{\eta}\right)^{3/4}\right. \\
&+&\left. \int_a^{v_s|\Delta\tau|}\frac{d\eta}{a}\left(\frac{a}{\eta}
\right)^{K_c+K_s-5/4} \right]  .
\ea
The requirement that $d(\bar{\lambda}_{\rm eff}a)/d\ell=0$ results in the
renormalization equations for the string parameters in Eqs. (\ref{RGfloppy}).


\begin{thebibliography}{999}

\bibitem{LeeRam} For a review see, P.~A.~Lee and T.~V.~Ramakrishnan,
Rev. Mod. Phys. {\bf 57}, 287 (1985).

\bibitem{Apel} W.~Apel and T.~M.~Rice, \prb {\bf 26}, 7063 (1982).

\bibitem{gs} T.~Giamarchi and H.~J.~Schulz, \prb {\bf 37}, 325  (1988).

\bibitem{Finkel} A.~M.~Finkel'stein, Z. Phys. B {\bf 56}, 189 (1984).

\bibitem{Castellani} C.~Castellani, C.~Di~Castro. P.~A.~Lee, M.~Ma,
S.~Sorella, and E.~Tabet, \prb {\bf 30}, 1596 (1984).

\bibitem{Punn} A.~Punnoose and A.~M.~Finkel'stein, Science {\bf 310},
289 (2005).

\bibitem{feldman} D.~E.~Feldman and Y.~Gefen, \prb {\bf 67}, 115337
(2003).

\bibitem{martin} P.~San-Jos$\acute{\rm e}$, F.~Guinea, and T.~Martin,
\prb {\bf 72}, 165427 (2005).

\bibitem{caroli}
C. Caroli, P.G. de Gennes and J. Matricon, Physics Lett. {\bf 9},
307 (1964).

\bibitem{stripereviews} For recent reviews of some of the experimental
evidence for the existence of meso-structures in the cuprates and their  
possible relation to the mechanism of high-temperature superconductivity see
E.~W.~Carlson, V.~J.~Emery, S.~A.~Kivelson and D.~Orgad in {\it ``The
Physics of Conventional and Unconventional Superconductors''}, Vol 2,
edited by K.~H.~Bennemann and J.~B.~Ketterson (Springer-Verlag 2004),
and S.~A.~Kivelson, I.~P.~Bindloss, E.~Fradkin, V.~Oganesyan,
J.~M.~Tranquada, A.~Kapitulnik and C.~Howald, Rev. Mod. Phys. {\bf 75},
1201 (2003).

\bibitem{1dreview}  For reviews, see V.~J.~Emery in {\it Highly Conducting
One-Dimensional Solids}, edited by J.~T.~Devreese, R.~P.~Evrard, and
V.~E.~van Doren, (Plenum, New York, 1979) p. 327, and T.~Giamarchi,
{\it Quantum Physics in One Dimension}, (Oxford University Press, Oxford,
2004).

\bibitem{jose}J.~V.~Jos$\acute{\rm e}$, L.~P.~Kadanoff, S.~Kirkpatrick, and
D.~R.~Nelson, \prb {\bf 16}, 1217 (1977).

\bibitem{invcomm} The relation $dK_s/d\ell=1/2 (dy/d\ell)$ holds up
to first order in ${\cal D}_b$ and second order in $y$, separately. The
terms of order $y{\cal D}_b$ in the flow equations do not obey this relation.
We expect terms of order $y^2{\cal D}_b^2$ in the expansion of $R_s$, which
were not taken into account here, to fix this problem.

\bibitem{string} D.~Orgad, \prb {\bf 69}, 205104 (2004).

\bibitem{elasticom} This is true also for the case where the electronic
density along the string, $n$, is kept fixed rather than the number of
particles. Under such conditions the total kinetic energy of the
electrons fluctuates as they flow in and out of the string in order
to compensate for the changes in the string length. The result is
an induced elastic term with $\sigma=E_F n$, where $E_F$ is the (fixed)
Fermi energy of the electrons.

\bibitem{schemecom} The result is sensitive to the regularization scheme
used in evaluating the average. Here and in the following we employ a
hard cutoff $1/a$ in the integral over wave-vectors and a hard cutoff
$\gamma/a$ in the frequency integrals.

\bibitem{2legdis} E.~Orignac and T.~Giamarchi, \prb {\bf 53}, R10453
(1996); \prb {\bf 56}, 7167 (1997).

\bibitem{2legbos} E.~Orignac and T.~Giamarchi, \prb {\bf 57}, 11713 (1998).

\bibitem{optimal} E.~Arrigoni and S.~A.~Kivelson, \prb {\bf 68}, 180503 (2003).
\end{thebibliography}
\end{document}